\documentclass[aps,jcp,superscriptaddress,amsmath,amssymb]{revtex4-2}

\usepackage{comment}
\usepackage{tikz}
\usetikzlibrary{trees}
\usetikzlibrary{decorations.pathmorphing}
\usetikzlibrary{decorations.markings}
\usepackage{tikz-cd}
\usetikzlibrary{arrows,arrows.meta}
\usetikzlibrary{decorations.pathreplacing,snakes}
\tikzset{
    photon/.style={decorate, decoration={snake,segment length=1.5mm}, draw=black},
    coulomb/.style={dotted},
    electron/.style={draw=black, postaction={decorate},
        decoration={markings,mark=at position .55 with {\arrow[draw=black]{>}}}}, 
    gluon/.style={decorate, draw=magenta,
        decoration={coil,amplitude=4pt, segment length=5pt}},
    boundelectron/.style={thick, double},
    transverse/.style={dashed},
    marrow/.style={decoration={markings,mark=at position 0.5 with {\arrow{#1}}}, postaction=decorate}    
}
\usepackage{graphicx}
\usepackage{dcolumn}
\usepackage{bm}
\usepackage{rotating}

\newcolumntype{.}{D{.}{.}{8}}

\usepackage[utf8]{inputenc}
\usepackage{physics}
\usepackage{amsmath, amssymb}
\usepackage{tabularx,booktabs,array,dcolumn,multirow}
\usepackage{setspace}
\usepackage{vmargin}
\usepackage{xcolor}
\usepackage{graphicx,psfrag,subfigure}

\usepackage{float}
\usepackage[all]{xy}

\usepackage{bm}
\usepackage{mathtools}
\usepackage{physics}
\usepackage[english]{babel}

\newcommand{\mr}[1]{\mathrm{#1}}
\newcommand{\cm}{cm$^{-1}$}

\def\Eh{E_\mr{h}}

\definecolor{carnelianred}{cmyk}{0, 0.864, 0.837, 0.274}
\definecolor{harvestgold}{cmyk}{0, 0.330, 0.981, 0.133}
\definecolor{patricksblue}{cmyk}{0.826, 0.644, 0, 0.525}
\definecolor{arany}{cmyk}{0, 0.295, 0.988, 0.003}
\definecolor{red}{cmyk}{0, 0.785, 0.785, 0.176}
\definecolor{blue}{cmyk}{0.680, 0.551, 0, 0.423}

\definecolor{gray}{rgb}{0.5,0.5,0.5}

\usepackage[unicode]{hyperref}
\usepackage{soul}
\hypersetup{
   unicode=true,          
   plainpages=false,
   colorlinks=true,       
   linkcolor=black,          
   linkcolor=blue,          
   citecolor=blue,        
   urlcolor=blue           
}

\usepackage{url}

\begin{document}
\noindent

\title{%
The Molecular Quantum \emph{electro-}Dynamics Research Group in Budapest
}

\author{Edit M\'atyus} 
\email{edit.matyus@ttk.elte.hu}
\affiliation{ELTE, Eötvös Loránd University, Institute of Chemistry, 
Pázmány Péter sétány 1/A, Budapest, H-1117, Hungary}

\date{\today}

\begin{abstract}
\noindent %
This article briefly overviews the scientific activities of the Molecular Quantum (electro-)Dynamics (MQD) Research Group in Budapest. Since its foundation in 2016, the MQD group has worked on molecular spectroscopy and molecular physics topics with primary applications and relevance to high-resolution and precision spectroscopy. \\

\noindent %

\vspace{0.5cm}
\noindent %
\emph{To appear in the `Hungarian Quantum Chemistry' volume of the Advances in Quantum Chemistry.}
\end{abstract}

\maketitle

The Molecular Quantum Dynamics research group was established in the Institute of Chemistry of ELTE, Eötvös Loránd University, in 2016 with funding from the Swiss National Science Foundation (SNF). The SNF funding was part of the first Swiss contribution to the development of the European Union's `New Member States', \emph{i.e.,} countries from the former Eastern block of Europe, including Hungary. 
The MQD research group was launched with two main research directions: (a) the development of the theoretical foundations and computations for precision spectroscopy beyond the Born-Oppenheimer and non-relativistic approximations and (b) the numerical solution of the rovibrational Schrödinger equation of floppy systems, with focus on molecular clusters and complexes. The two main research themes were complemented with a smaller, more philosophically inclined direction (c) aimed at the development of (classical) molecular concepts in quantum mechanics, \emph{e.g.,} the molecular structure and molecular shape in quantum mechanics and, in general, better understanding the role of decoherence in molecular systems \cite{Ma19review,MuMaRe19,MaCa21}.

The precision spectroscopy direction (a) has gained additional momentum through funding by an ERC Starting Grant. The ERC Starting Grant project, POLYQUANT: Theoretical developments for precision spectroscopy of polyatomic and polyelectronic molecules, is about theoretical and computational developments of non-adiabatic, relativistic, and quantum-electrodynamics (QED) effects in molecular systems. 
Most recently, the relativistic QED research direction resulting from one of the ERC subprojects was reinforced with major local funding through the Momentum Programme of the Hungarian Academy of Sciences, and hence, the `electro-' extension in the name, Molecular Quantum electro-Dynamics (MQeD) Research Group. 

\subsection{The beginnings}
During 2016-17, I thought the MQD group, including myself, might be a single-element group in the Institute of Chemistry at ELTE. I applied for the permission of the Institute's Board to register the official name `Molecular Quantum Dynamics Research Group' only a year later, after Dr Gustavo Avila joined me.
The early years were spent with rovibrational computations on floppy molecular complexes
and contemplating between variational (pre-BO) and perturbative (non-adiabatic mass) non-Born--Oppenheimer approaches to achieve better agreement with the high-resolution experimental spectra for small systems.

Regarding the floppy molecular systems, I continued to work on the methane-water complex (that I knew from my post-doc time in Cambridge) \cite{SaCsAlWaMa16,SaCsMa17,dimers,DaAvMa21mw}. In this series of work, we have finally solved the high-resolution rovibrational spectrum of methane-water recorded in the 1990s \cite{DoSa94}. 
For small molecules and ultra-high precision, I was looking for meaningful handles to take off with relativistic corrections. At some point during 2017, I received advice from a good colleague. For a start, I should not worry about a Dirac-relativistic approach---which was clearly a major theoretical challenge---just start off with the computation of the Breit-Pauli Hamiltonian expectation value to get the leading-order relativistic corrections for compounds of the light elements.
In the meanwhile, I was talking to (particle) physics colleagues in Budapest, telling them that I was looking for a fully covariant equation that I could use for molecular computations. I have learned from them about lattice-based approaches (successfully used in quantum chromodynamics, QCD). Lattice QCD has a modern, well-readable literature, with accessible tutorials and textbooks, \emph{e.g.,} \cite{GaLaBook09}. Nevertheless, the theoretical framework was very different (even for a possible QED variant, which has conceptual and technical problems) from the quantum chemistry and (exact) quantum dynamics approaches, \emph{i.e.,} solving a differential equation, that I knew to be very well applicable for numerically computing molecular energies.
The path-integral tunnelling splitting computations that I have learned about in Cambridge \cite{MaWaAl16,MaAl16} were a bit closer to the lattice QCD philosophy (both operating with the Lagrangian, which is easier to formulate for QED and QCD than the Hamiltonian), but I had in mind computing (ground and excited) rovibronic states of small molecules for high-resolution and precision spectroscopy applications. So, instead, I continued searching for an eigenvalue-like equation.
During this time, I came across the Bethe-Salpeter equation, but it was not straightforward at this stage to use it in practical computations for atoms and molecules.

\vspace{0.5cm}
\subsection{Exact quantum dynamics developments and computational applications}
Dr. Gustavo Avila was the first post-doctoral co-worker to join the MQD Group. With him, we continued thinking about the numerical solution of the rovibrational Schrödinger equation of floppy molecular systems of higher dimensionality. His former work with Tucker Carrington on the grid pruning approach, named after the Russian mathematician Smolyak \cite{AvCa09,AvCa11,AvCa11b}, was truly remarkable. The Smolyak scheme made it possible to attenuate the exponential growth of the vibrational integration grid with the number of vibrational degrees of freedom. Still, it was applicable only for semi-rigid systems. I was wondering whether it could be used for floppy systems as well. What we did was a simple combination of two complementary approaches [G. Avila and T. Carrington, Jr., J. Chem. Phys. 131, 174103 (2009)] and [E. Mátyus, G. Czakó, and A. G. Császár, J. Chem. Phys. 130, 134112 (2009)] and was written up in the paper [Toward breaking the curse of dimensionality in (ro)vibrational computations of molecular systems with multiple large-amplitude motions, G. Avila and E. Mátyus, J. Chem. Phys. 150, 174107 (2019)].
The idea was very simple: we used the efficient basis and grid truncation for the semi-rigid part of the system, whereas we continued using a direct product representation for the more difficult floppy part. The numerical kinetic energy operator approach \cite{MaCzCs09} allowed us to use efficient (even numerically defined, \emph{vide infra}) coordinates, which is a prerequisite for efficient basis and grid truncation. This idea was first used for the complexes of methane, for the weakly bound floppy CH$_4\cdot$Ar \cite{AvMa19} and for the more strongly bound CH$_4\cdot$F$^-$ \cite{AvMa19b,PaTaAvMaCz23}. Later developments by Gustavo Avila made it possible to reach the lowest-energy, predissociative vibrational states of the methane moiety within the complex (to be published).

In later years, this approach was extended and adapted to converging the vibrational states of the formic acid (HCOOH) molecule (9D) \cite{AvDaMa23}, partly motivated by recent experiments in Martin Suhm's group in Göttingen \cite{NeSu20,NeSi21}. Arman Nejad's visits to Budapest,---Arman's PhD project included measuring the vibrational infrared and Raman spectra of the formic acid (and isotopologues) in a supersonic jet---, led to further variational computations of better-converged vibrational energies and the development of electric dipole and polarizability surfaces \cite{AvDaMa23}.

Alberto Mart\'in Santa Dar\'ia joined the MQD group in 2018 as a PhD student and did beautiful work on rovibrational computations, including HCOOH. We defined (path-following) curvilinear normal coordinates for the small-amplitude vibrations of HCOOH, which was implemented in the numerical kinetic energy operator approach \cite{MaCzCs09}. The basis and the Smolyak grid truncation could be efficiently used for this coordinate definition. At the same time, the computation fully accounted for the floppy torsional degree of freedom without any truncation. We computed the infrared and Raman (approximate) vibrational spectra. The well-converged vibrational states (to be published) will be used in subsequent rovibrational computations. A combination of the electric dipole and polarizability surfaces \cite{AvDaMa23} will be used to compute the rovibrational infrared and Raman transition energies and transition moments.

Most recently, Dr. Ayaki Sunaga joined the group as an MSCA Post-Doctoral Fellow. He extended the formic acid (9D) computations to the methanol (CH$_3$OH) molecule (12D) using path-following curvilinear normal coordinates. This setup allowed him to converge the vibrational intervals to 0.5~\cm\ up to the first overtone of the CO stretch, at 2200~\cm\ beyond the zero-point vibrational energy \cite{SuAvMa24}. The lowest-energy states are in excellent agreement with vibrational band origins derived from the experiments. Still, for higher-energy combination bands, larger deviations (20-30~\cm) have been attributed to imperfections of the potential energy surface (PES), which is now under improvement.

Most recently, motivated by recent developments in quantum logic spectroscopy (a novel spectroscopic technique exploiting quantum logic \cite{ScRoLaItBeWi05}) towards the molecular regime \cite{SiMeNaHeWi20,ChCoKuLiHaPlFoDiLeLe20,SiWi23}, we have extended the rovibrational modelling of polyatomic systems with accounting for the hyperfine and Zeeman effects \cite{AvSuKoMa24} to leading order in the fine-structure constant and using a non-relativistic reference. The first applications are reported for the H$_3^+$ molecular ion \cite{AvSuKoMa24}, for which the numerical error control of both the electronic and the nuclear quantum mechanical treatment can be provided to high precision, and potentially, further non-adiabatic, relativistic, and QED corrections can be accounted for. 

In traditional quantum chemistry and quantum dynamics, the cancellation of the numerical error and some physical effects is powerfully exploited during the modelling of quantum mechanical motions taking place at different scales, \emph{i.e.,} the electronic vs. vibrational. vs. rotational motions~\cite{PrMH81}. We anticipate a similar cancellation of error given the different, \emph{i.e.,} the proton spin vs. molecular rotation and vibration, scales of motions. Nevertheless, at a sufficiently high experimental resolution, which can be anticipated from the recent adaptation of modern quantum technology techniques to molecular ion experiments, an explicit account of further effects and couplings will be necessary, which will allow us to probe and extend the current boundaries of molecular quantum mechanics computations.

\subsection{Non-adiabatic, relativistic, and QED corrections}
During the initial years of the MQD group, variational pre-Born--Oppenheimer developments characterized the main activities \cite{Ma13,Ma19review,FeMa19EF}, as a continuation of my post-doctoral research \cite{MaHuMuRe11a,MaHuMuRe11b,MaRe12} as an ETH Fellow in Zürich, and partly in later collaboration with the Reiher group, where Benjamin Simmen, and then, Andrea Muolo were PhD students during that time \cite{SiMaRe13,SiMaRe14,SiMaRe15,MuMaRe18,MuMaRe19}. We have worked on projection techniques of floating explicitly correlated Gaussians \cite{MuMaRe18}, which resulted in several orders of magnitude improvement in the five-particle non-relativistic (rotational) energy of H$_3^+$, but was still ca.~one order of magnitude less accurate that could be computed from a Born--Oppenheimer (BO) based approach with perturbative post-BO corrections. An early work about solving the two-electron Dirac--Coulomb equation with an explicitly correlated Gaussian basis also resulted from this collaboration \cite{SiMaRe15}.

As an alternative to fully abandoning the BO approximation, I have studied a perturbative approach to correcting for the post-BO effects \cite{Ma18nonad,Ma18he2p}. The most interesting problem in this direction was the nuclear kinetic-energy correction term, which led to the (formerly, partly empirically known) non-adiabatic (effective) mass concept of the rotating-vibrating nuclei. Regarding the development of the formal theoretical background, I have benefited from work with Stefan Teufel~\cite{MaTe19}, which led to the formulation of the non-adiabatic perturbative corrections not only for single electronic states but also for coupled electronic states, and made it possible to perturbatively correct for all outlying electronic states beyond the explicitly coupled electronic subspace. Non-adiabatic mass corrections for the coupled EF-GK-HH electronic manifold (subspace) of H$_2$ were computed in a pilot study \cite{MaFe22nad}.

\begin{figure}
  \includegraphics[width=10cm]{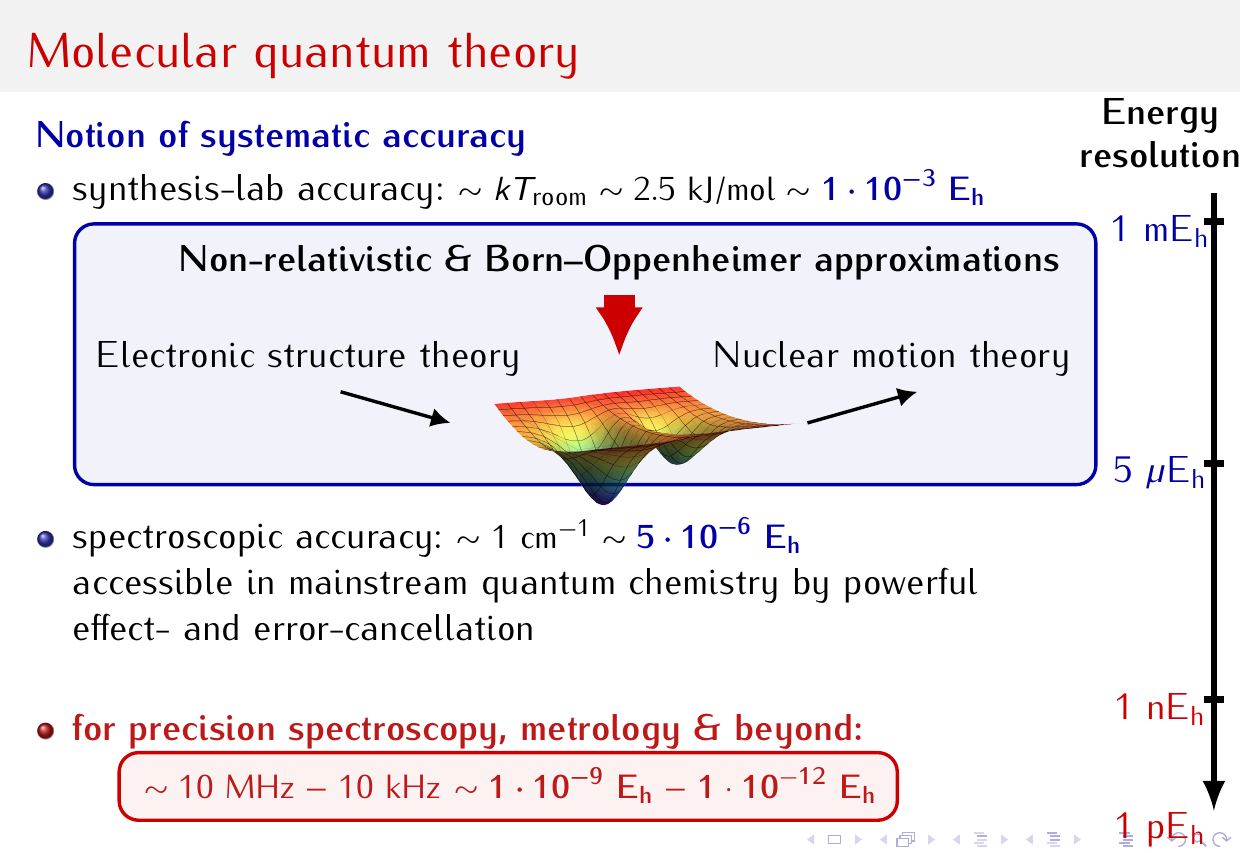}
  \caption{%
    Orientation chart regarding the typical, spectroscopic and metrologic, energy resolution relevant to activities of the Molecular Quantum electro-Dynamics Research Group. 
    \label{fig:resolution}
  }
\end{figure}

We computed the non-adiabatic mass correction for a series of diatomic systems \cite{Ma18nonad,Ma18he2p,FeMa19HH,FeKoMa20}, which could be characterized by a single-electronic state. The perturbative post-BO corrections (the diagonal BO correction and the non-adiabatic mass correction) were essential to obtain 
quantitative agreement with high-resolution spectroscopy experiments, but it was also clear that good agreement with the experiment can only be achieved if the relativistic and QED corrections
are explicitly computed. For the small and light systems, the non-adiabatic, relativistic, and leading-order QED corrections to the rovibrational intervals were about the same order of magnitude, and their precise magnitude and sign have changed from interval to interval and from system to system \cite{Ma18he2p,FeKoMa20,FeMa19HH,FeMa19EF}, it was impossible to reasonably rely on cancelling effects at this range and energy resolution (Fig.~\ref{fig:resolution}).

Regarding the relativistic and QED corrections, I was contemplating different approaches. I have learned about four-component relativistic approaches from Markus Reiher \cite{ReWoBook15} (see also the related textbooks Refs.~\cite{DyFaBook07,GrBook07,LiBook11}), successful in predicting and rationalizing the chemical and physical properties of heavy- and superheavy-elements and their compounds \cite{Py12,ScSmPy20,PaElBoKaSc17,ScPaPuBo15,QuSkGr98,GrQu22,bertha20,dirac20}. At the same time, I have learned about a fully perturbation theory-based approach, starting with the Breit-Pauli Hamiltonian expectation values, appended with further corrections for the leading and higher-order QED corrections with fabulous success in the precision spectroscopy of light elements \cite{KoPiLaPrJePa11,PrCeKoLaJeSz10,WaYa18,PuKoCzPa16,PuKoCzPa19}. The two approaches were very different, and I could not reconcile them then (2018). I was hesitating about which direction to pursue. 

In 2018, Dávid Ferenc joined the group (first as an MSc, and then) as a PhD student. With him, we started to implement the matrix elements for the Breit-Pauli (BP) Hamiltonian with explicitly correlated Gaussian functions. For debugging the code, there were two parallel implementations, and one of them was more general and already enabled for the extra (Cartesian) indices, which were necessary for solving the two-electron Dirac equation. One of the first applications of the final working implementation, appended with an initial regularization scheme that was found to be essential for the singular BP terms and a (explicitly correlated) Gaussian basis set, and with help from Vladimir Korobov with (estimating and) computing the Bethe logarithm for the ion core, our work on the rovibrational intervals of $^4$He$_2^+$ (X~$^2\Sigma_\text{u}^+$) including non-adiabatic, relativistic and QED corrections was published in the Physical Review Letters in 2020. 

In 2020, Dr. Péter Jeszenszki joined the group, and Dávid was already a well-versed, second-year PhD student. At this point, also with funding from the ERC, I decided to finally take off with the development of a Dirac-relativistic approach, ultimately still aiming at applications for precision (atomic and molecular) spectroscopy. I knew the old wisdom: when you hesitate about which direction to pursue, you should do both. So, we have taken off with the more risky variational relativistic direction, in addition to the better-established perturbation theory approach (for compounds of light elements). 

The first numerical variational relativistic results of two-electron systems \cite{JeFeMa21}, after extensive debugging and testing, were puzzling due to the `large' deviation (on the typical energy scale of precision spectroscopy) from the leading-order perturbation theory results already for the helium ground state with the (small) $Z=2$ nuclear charge number.  By this time, we already had experience with the numerically stable and well-converged computation of the Breit-Pauli expectation value with explicitly correlated Gaussian basis sets (\emph{vide infra} about the essential regularization schemes for the perturbative route), which we used to check the variational results. Soon later, we realized that the variational relativistic computation (of course) automatically included higher-order relativistic contributions, which were visible and significant already for the helium atom (and even for the H$_2$ molecule), at the typical energy scale of precision spectroscopy (Fig.\ref{fig:resolution}). 

After the implementation of the no-pair Dirac--Coulomb model \cite{JeFeMa21,JeFeMa22} with the empirical cutting and the more rigorous (but technically more problematic) complex coordinate rotation (CCR) positive-energy projector (adapted from pioneering work by Bylicki, Pestka, and Karwowski \cite{ByPeKa08}), we finally completed also the Breit term \cite{FeJeMa22,FeJeMa22b}. Further extensive testing followed, which led to a better understanding of the numerical behaviour of the results; and we have finally observed an excellent numerical agreement \cite{FeJeMa22b} of the $\alpha$ fine-structure dependence of the variational relativistic energy in comparison with the perturbative relativistic results. 
In the literature, we could identify some of the higher-order perturbative relativistic corrections (separately from the other terms at the specific $\alpha$ order), which further confirmed our variational results (and helped develop a better understanding). 
In cross-checking our variational no-pair relativistic results with perturbation theory using a non-relativistic reference (the so-called `nrQED' scheme), we found extremely useful Joseph Sucher's PhD thesis: \emph{Energy Levels of the Two-Electron Atom, to Order $\alpha^3$ Rydberg} (Columbia University, 1958) \cite{SuPhD58}. 

By 2023, Péter Jeszenszki pinpointed the triplet contributions to the dominant singlet ground states of the helium atom and the hydrogen molecule \cite{JeMa23}, for which the $\alpha$ fine-structure dependence was again in excellent agreement with the perturbation theory literature.

Meanwhile, Dávid Ferenc has worked on the pre-Born--Oppenheimer generalization of our newly developed two-spin-1/2-fermion relativistic formalism and computer program. This was not a single-run success; we have repeatedly revisited the two-body relativistic pre-BO problem. In the end, the careful study of Salpeter's original paper \cite{Sa52} led to an unambiguous and well-defined approach with high-precision numerical results. Again, the $\alpha$ fine-structure-constant dependence of the variational relativistic energy was in excellent agreement with the perturbation theory result of the respective $\alpha$ order \cite{FuMa54}. For the first time, we saw the logarithmic term of $\alpha^4\ln\alpha\Eh$ order in the $\alpha$ series expansion of the variational relativistic energy \cite{FeMa23}, in full agreement with the available perturbation theory literature \cite{KhMiYe93,Zh96}. 

During 2024, Ádám Nonn elaborated on the linear algebra subroutines and parallelized their in-house, increased (quadruple) precision version. This development allowed him to better converge the no-pair Dirac-Coulomb(-Breit) (DC(B)) energy of the two-electron helium atom approaching the parts-per-trillion (ppt) relative precision. As a result, he numerically determined the $\alpha^4\ln\alpha$ prefactor from the variational computations \cite{NoMaMa24}, which was (again) in excellent agreement with the value known from perturbation theory, thereby demonstrating the consistency of our newly developed variational approach.

By 2023-2024, we have demonstrated that the variational relativistic computations can be converged to high precision (using the LS coupling scheme with an explicitly correlated Gaussian basis set) relevant for precision spectroscopy of compounds of light elements. The fine-structure dependence of the variational energy was demonstrated to be in excellent agreement with the available results of the state-of-the-art nrQED scheme, which has so far provided the most stringent tests of experimental molecular spectroscopy results. Furthermore, our variational relativistic approach was defined based on the rigorous theoretical foundation of relativistic quantum electrodynamics and the Bethe--Salpeter equation \cite{Fe49a,Fe49b,SaBe51,Sa52,SuPhD58,DoKr74}, but allowed us to use a relativistic reference state, which included a partial resummation in $(Z\alpha)^n$ known to be qualitatively important for higher-$Z$ systems.

\subsection{Molecular relativistic quantum electrodynamics}
These results and observations led to the definition of a new research programme \cite{MaFeJeMa23} for the development of the theoretical formulation and for the implementation of a practical, numerical approach, which uses the correlated, variational relativistic state as a reference for relativistic QED computations. 
The first exploratory steps in this direction \cite{MaMa24} were taken with Dr. Ádám Margócsy, who started to work on the self-energy correction, which is the largest QED correction in light systems and which requires renormalization. 
Furthermore, Ádám Nonn \cite{NoMaMa24} started to work on the `C$\times$C' crossed Coulomb-photon correction, which is the simplest crossed-photon correction to the electron-electron (or, in general, the two-spin-1/2 fermion) interaction. In addition, the leading-order perturbative correction to the relativistic reference for the retarded photon exchange (transverse photon exchange in the Coulomb gauge) has also been formulated \cite{MaFeJeMa23,MaMa24} based on the old literature and by considering the practical numerical framework. The ongoing progress along all lines is promising. 

Last but not least, defining the positive- and negative-energy projectors was an important element for the numerical computation of QED corrections to a correlated relativistic reference. This topic was touched upon already during the early no-pair DC(B) computations \cite{JeFeMa21,JeFeMa22}, but neither the empirical energy cutting nor the CCR projection technique provided a general route for constructing projectors to any of the four two-fermion ($++,+-,-+,--$) subspaces. Any progress became possible after the rigorous formulation of an elementary problem: the faithful basis representation of one-particle operators over non-separable two-particle basis states \cite{NoMaMa24,HoJeMa24}. This development was essential for working on the basis-set representation of the various QED corrections and is already extensively used in the C$\times$C and transverse interaction implementations (to be published).

\paragraph{Perturbative regularization approaches to the leading-order relativistic energy correction}
In parallel with the variational relativistic developments (for the moment focusing on two-spin-1/2 fermion systems), we developed practical regularization schemes for the precise evaluation of the Breit-Pauli Hamiltonian expectation value using (explicitly-correlated) Gaussian basis sets, applicable to molecular systems with more than two electrons. 
The early work \cite{JeIrFeMa22} elaborated on the integral transformation (IT) technique \cite{PaCeKo05}, which we used already in Ref.~\citenum{FeKoMa20}. The IT approach uses a cutoff, which had to be adjusted manually. This procedure was ambiguous and prone to error, especially for the mass-velocity correction term. Most recently, Balázs Rácsai has implemented a numerical Drachmannization approach \cite{RaFeMaMa24}, which we found to be more robust than the IT approach and which has opened the route to the automated computation of the leading-order (BP) relativistic correction for molecular systems at hundreds or thousands of nuclear configurations of various electronic states.

\subsection{Present and future}
Almost 10 years after its foundation, the main goals of the MQD group have not changed. It works on theoretical and methodological developments related to high-resolution and precision (molecular and atomic) spectroscopy experiments. 

For the next several years, I think the core activities will focus on the fundamental theoretical developments of relativistic QED for small atomic and molecular systems. One of my near-future goals is to test the non-radiative part of nrQED of triplet helium. The radiative and non-radiative nrQED corrections have been derived and computed up to $\alpha^5\Eh$ order for triplet helium (singlet helium appears to be limited to $\alpha^4\Eh$ for the foreseeable future). Still, there has been a significant deviation of theory and experiment, known as the triplet helium puzzle, already for five years \cite{PaYePa20,PaYePa21,YePaPa23,ClJaScAgScMe21}.

I also aim to remain open for experimentally motivated computations, primarily for high-resolution and precision spectroscopy. 
We are continuously improving our ECG-based non-adiabatic, relativistic and QED implementation, and it will become amenable to larger-scale computational applications. For this purpose, we elaborate on the non-relativistic optimization scheme, both in terms of precision and in terms of system size; we aim to turn the in-principle established approaches numerically more robust, for which an important step was the development of the numerical drachmannization scheme \cite{JeIrFeMa22,RaFeMaMa24} and the implementation of the Bethe logarithm \cite{FeMa23bethe}. 

I would like to have computed instead of estimated error bars for the computed results. In this respect, the development of a lower bound theory for the (non-relativistic) energy is of utmost importance, which has seen major progress in recent years through collaboration with Eli Pollak's group \cite{IrJeMaMaRoPo22,RoJeMaPo23}, and its practical applications will follow in the future.

For small polyatomic molecular systems, in which the rovibrational energies (for a given PES) can be converged to high precision, we have recently developed a computational procedure to account for the hyperfine-Zeeman effect in the high-resolution rovibrational spectrum \cite{AvSuKoMa24}. I am curious to see whether the best quantum chemistry and quantum dynamics models are sufficient for modelling, predicting, and tracking small magnetic effects in the high-resolution spectrum, which may be even for quantum technological relevance in the future.

\subsection{Members and Visitors of the MQD Research Group since 2016}
I am grateful to many colleagues, co-workers, and research students. Thanks to them, the MQD Group, initiated in 2016, did not end up as a single-element group. The work I have briefly reviewed in this article was made possible by the hard work and enthusiasm of several MQD group members over the past several years. 
Part of the MQD group efforts was arranged to train the next generation of students and co-workers. Several PhD and post-doctoral group members were involved in co-supervising the younger students.
At Hungarian higher education institutions, research studentship has a long tradition, and the students are encouraged to join a research group and work on a small research project in parallel with their university studies. The research students are also encouraged to participate in the yearly institutional research student competition (called `TDK') and the biannual national-level `OTDK' competition. Certainly, not every excellent student participates in the (O)TDK. Still, this student program often helps the students gain practical experience in research and deepen their knowledge of a specific problem. The TDK programme has a strong mentoring element that contributes to high-quality education (1:1 student-teacher ratio), and it also requires major participation of research groups in the Institute's educational activity.

\vspace{0.55cm}
\noindent%
\textbf{Research students} \\
Several research students have been working in the MQD group. The following list includes the students whose work resulted in an original research article or a TDK/BSc/MSc thesis.
\begin{itemize}
  \item 
    {Ádám Nonn} (Chemistry MSc), 09/2023-- 
  \item
    {Dezs\H{o} Palik} (Chemistry BSc), Summer 2024-- 
 \item
    {Balázs Rácsai} (Chemistry BSc), Summer 2023-- 
  \item
    {Eszter Saly} (Chemistry BSc, MSc), Summer 2021-- 
  \item
    {Péter Hollósy} (Physics BSc, MSc), Summer 2021-- 
  \item
    {Robbie Ireland} (Erasmus+ Trainee Scholarship from Glasgow University), 09/2020--07/2021 
  \item
    {Dávid Ferenc} Spring 2018 (then, as PhD student until 2022) 
\end{itemize}

\vspace{0.45cm}
\noindent %
\textbf{PhD students}
\begin{itemize}
  \item 
    {Alberto Mart\'in Santa Dar\'ia,} 8/2018--9/2022 
  \item
    {Dávid Ferenc,} 7/2018--9/2023 
\end{itemize}

\vspace{0.45cm}
\noindent %
\textbf{Post-doctoral researchers} 
\begin{itemize}
  \item 
    {Dr. László Biró,} 2024 
  \item
    {Dr. Ayaki Sunaga,} 9/2023-- 
  \item
    {Dr. Ádám Margócsy,} 7/2022-- 
  \item
    {Dr. Miklós Rontó,} 6/2022--8/2022 
  \item
    {Dr. Mykhaylo Khoma,} 2020 
  \item
    {Dr. Péter Jeszenszki,} 1/2020-- 
  \item
    {Dr. István Hornyák,} 3/2020--6/2022 
  \item
    {Dr. Gustavo Avila,} 8/2017-- 
  \item
    {Dr. B\'ela Szekeres} (50\% ), 10/2016--2/2018 
\end{itemize}

\vspace{0.65cm}
\textbf{Visitors of the MQD Research Group}
\begin{itemize}
  \item
    {{Prof. Dr. David Lauvergnat} (Université Paris-Saclay, CNRS): May 2024 (2 days)}
  \item
    {{Prof. Dr. Ulrich Jentschura} (Missouri S\&T): November 2023 (2 days)} 
  \item
    {{PD. Dr. Zoltán Harman} (MPI Heidelberg): October 2023 (2 days)} 
  \item
    {{Prof. Dr. Albert Bartók-Pártay} (Warwick): May 2023 (3 days)} 
  \item
    {{Prof. Dr. Stanislav Komorovsk\'y} (Bratislava): April 2023 (1 day)} 
  \item
    {{Dr. Alberto Mart\'in Santa Dar\'ia} (Salamanca): April 2023 (1 week)} 
  \item
    {{Dr. Arman Nejad} (Göttingen): September 2022, March 2023 (2 weeks)} 
  \item
    {{Prof. Markus Reiher} (Zürich): November 2019 (2 days)} 
  \item
    {{Prof. Vladimir Korobov} (Dubna): October 2019 (2 days)} 
  \item
    {{Prof. Stefan Teufel} (Tübingen): September 2019 (2 days)} 
  \item
    {{Prof. Patrick Cassam-Chenai} (Nice): August 2019 (2 weeks)} 
\end{itemize}

\vspace{0.75cm}
\noindent %
\emph{Acknowledgement:} The Author thanks the financial support of the Swiss National Science Foundation (No. IZ11Z0\_166525), the European Research Council (No. 851421), the Hungarian National Research and Development Office (FK~142869), and the Momentum Programme of the Hungarian Academy of Sciences. The Hungarian HPC Infrastructure (NIIF Institute, Hungary) is acknowledged for allocating computer time. The Author also thanks the hospitality of the DFG BENCh Training School in Göttingen and the Institute of Molecular Physical Sciences at ETH Zürich and acknowledges participation in the COST-MOLIM Network.

\vspace{0.75cm}
\noindent
\emph{Biography of the Author:} Edit Mátyus attended the science class of the ELTE Apáczai Grammar School, then studied chemistry at ELTE and defended her doctoral dissertation in November 2009. After post-doctoral studies at ETH Zürich and the University of Cambridge, she became an assistant and then an associate professor at her alma mater. Dr Mátyus has been awarded the Dirac Medal (2021) of WATOC and the Annual Medal (2023) of the International Academy of Quantum Molecular Science.

\vspace{0.5cm}

\end{document}